\documentclass[runningheads]{llncs}
\usepackage{graphicx}
\usepackage{draftfigure}
\usepackage{amsmath}
\usepackage[numbers,sort&compress]{natbib}
\usepackage{tabularx}
\usepackage{hyperref} 
\usepackage{caption}
\usepackage{subcaption}
\usepackage{comment}
\usepackage{color}
\usepackage{enumitem}

\newcommand{\zinspection}{{Z-Inspection\textsuperscript{\tiny\textregistered}}}
\newcommand{\covid}{{COVID-19}}

\begin{document}
\title{Using Sentence Embeddings and Semantic Similarity for Seeking Consensus when Assessing Trustworthy AI}
\titlerunning{Using Semantic Similarity for Seeking Consensus}
\author{Dennis Vetter\inst{1}\orcidID{0000-0002-5977-5535} \and
Jesmin Jahan Tithi \inst{2} \and
Magnus Westerlund \inst{3} \and
Roberto V. Zicari \inst{3,4} \and
Gemma Roig\inst{1}\orcidID{0000-0002-6439-8076}}
% %
\authorrunning{D. Vetter et al.}
\institute{Goethe University Frankfurt, 60629 Frankfurt am Main, Germany
\email{vetter@em.uni-frankfurt.de} \;\; \email{roig@cs.uni-frankfurt.de}\and 
Intel Labs, Santa Clara, CA 95054, United States \and
Arcada University of Applied Sciences, 00550 Helsinki, Finland \and
Seoul National University, Seoul 08826, South Korea
}
\maketitle              % typeset the header of the contribution
\begin{abstract}
Assessing the trustworthiness of artificial intelligence systems requires knowledge from many different disciplines. These disciplines do not necessarily share concepts between them and might use words with different meanings, or even use the same words differently. Additionally, experts from different disciplines might not be aware of specialized terms readily used in other disciplines. Therefore, a core challenge of the assessment process is to identify when experts from different disciplines talk about the same problem but use different terminologies. In other words, the problem is to group problem descriptions (a.k.a. issues) with the same semantic meaning but described using slightly different terminologies.

In this work, we show how we employed recent advances in natural language processing, namely sentence embeddings and semantic textual similarity, to support this identification process and to bridge communication gaps in interdisciplinary teams of experts assessing the trustworthiness of an artificial intelligence system used in healthcare.
 
\end{abstract}
\section{Introduction}
The design, development and implementation of artificial intelligence (AI) systems requires knowledge from many different disciplines to be successful. Therefore, the teams involved in AI projects are often interdisciplinary to provide knowledge of all the relevant areas. Each area of expertise comes with its own specialized language, terms, definitions and jargon that can make communication between experts from different fields challenging, as they do not necessarily share the same concepts and may use the same words to mean something different \cite{whittlestone_ethical_2019}. Additionally, often time, people from one field might not be familiar with specialized terms used in another field. For example, an AI engineer might know the meaning of the terms ``precision and recall" whereas a healthcare professional may know the word ``prognosis" which the AI engineer might not know about.

One practical example where the interdisciplinary nature of communication shows up is the case where a team of interdisciplinary experts assesses an AI system for its trustworthiness \cite{zicari_z-inspection_2021,whittlestone_ethical_2019,ai_hleg_high-level_expert_group_on_artificial_intelligence_ethics_2019,ai_hleg_high-level_expert_group_on_artificial_intelligence_assessment_2020}. The stakeholders performing the assessment need to be aware of possible differences in the meaning of specialized terms so that they can understand each other properly. This requires them to cooperate with each other to work on a common vocabulary \cite{zicari_z-inspection_2021,whittlestone_ethical_2019}.

In this paper, we show how recent advances in the AI domain of natural language processing (NLP) can be used to support this process.
% use-case
Concretely, we apply it in the assessment of the trustworthiness of an AI system developed to evaluate the degree of lung damage in \covid~patients from their chest X-ray (CXR) images. Italian researchers developed the AI system in early 2020 to support the radiologists of a local hospital during the drastically rising cases of \covid~that overwhelmed the hospital system \cite{signoroni_bs-net_2021}. The goal of the system was to provide the radiologists with a qualified second opinion so they can work more confidently, faster, and with fewer mistakes.

\begin{figure}
    \includegraphics[width=\textwidth]{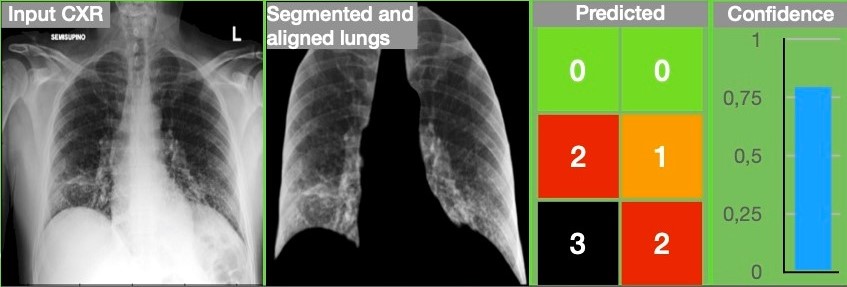}
    \caption{Schematic overview of the AI solution with the sub-tasks segmentation, alignment and Brixia score estimation. For the Brixia score, the lung is separated into 6 regions and each is rated with a number from 0 (no damage) to 3 (high damage) \cite{borghesi_chest_2020}. Image modified from \cite{signoroni_bs-net_2021}.}
    \label{fig:bs-net}
\end{figure}

The assessed AI system consists of multiple neural networks, one for each of the following sub-tasks: (1) segmentation of the CXR image into lung and background, (2) alignment of the image, and (3) estimation of the semi-quantitative Brixia score. For the Brixia score, the lung is separated into six regions and each region is assigned a number between 0 (no damage) and 3 (highly damaged). This separation into different areas and scoring based on a pre-defined set of values allows for efficient communication between radiologists \cite{borghesi_chest_2020}. A schematic view of the tasks performed by the AI system is given in Fig.~\ref{fig:bs-net}.

To train the networks, the researchers collected a large dataset of CXR images and annotations by either one radiologist (used for training) or the consensus of multiple radiologists (used for evaluation). Their results show that the AI system is performing equally well as an average human radiologist \cite{signoroni_bs-net_2021}.

% assessment process
The assessment of the above AI system \cite{signoroni_bs-net_2021} used the \zinspection~process described by Zicari et al. \cite{zicari_z-inspection_2021}, which is a holistic approach and includes participation of the entire community of key stakeholders. For assessing trustworthiness, \zinspection~builds on the \textit{Ethics Guidelines for Trustworthy AI} by the European Commission's High-Level Expert Group on AI with the four ethical principles of (i) respect for human autonomy, (ii) prevention of harm, (iii) fairness, and (iv) explicability, which are implemented through the seven key requirements of (1) human agency and oversight, (2) technical robustness and safety, (3) privacy and data governance, (4) transparency, (5) diversity, non-discrimination and fairness, (6) societal and environmental well-being, and (7) accountability \cite{ai_hleg_high-level_expert_group_on_artificial_intelligence_ethics_2019}.
%The results of the complete assessment are available at [] %\href{https://futurium.ec.europa.eu/en/european-ai-alliance/document/ai-hleg-assessment-list-trustworthy-artificial-intelligence-altai}, \href{https://robwortham.com/trustworthy-ai-a-new-assessment-tool-from-altai/}.

\begin{figure}[tbph]
    \centering
    \includegraphics[width=0.99\textwidth]{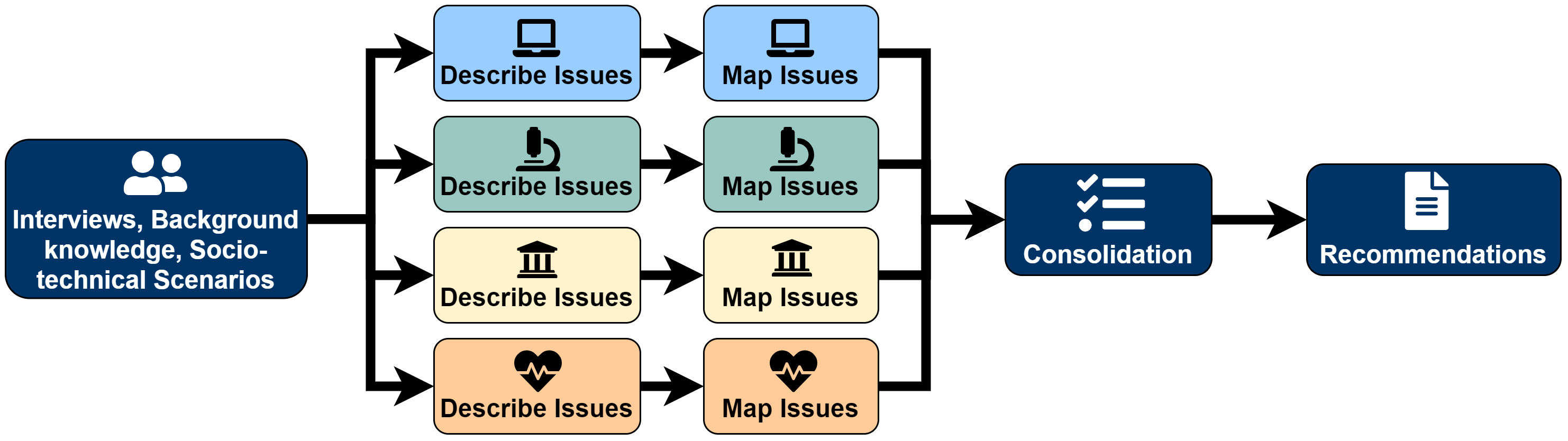}
    \caption{Schematic illustration of the mapping process. First step is to build a common knowledge base to develop socio-technical scenarios. Then the group is separated into WGs, according to the different backgrounds. The results of the WGs are combined in the consolidation step, based on which recommendations to the stakeholders are made. Adapted from \cite{zicari_z-inspection_2021}.}
    \label{fig:assessment_flow}
\end{figure}

% mapping
Part of the assessment is to use socio-technical scenarios \cite{leikas_ethical_2019,lucivero_ethical_2016} to identify different potential issues (ethical, legal, technical, etc) with the system, based on interviews with the whole team, the developers, other stakeholders, and additional materials such as academic papers, source code, datasets that are available. To achieve a disciplinary depth, the group of stakeholders is split into working groups (WGs) according to the different backgrounds of the participants. Each of these WGs then describes what potential issues/problems/tensions (conflicts between two or more desirable goals) they see with the system.
This is followed by the \textit{mapping} step, where the issues are structured and connected to the ethical principles and key requirements that they are conflicting with \cite{zicari_z-inspection_2021}. The goal of this mapping step is to have a description of the issues in ``structural ethical terms'' \cite{brusseau_what_2020}.
The output of this mapping is then used for \textit{consolidation} where a group-based consensus is reached regarding which issues can be combined and which issues are redundant. This consolidation allows to distill the most critical issues identified about the system; the consolidated statement is then reported to the system's developers and stakeholders, along with recommendations on possible steps to mitigate the issues or lower their impact.
A schematic illustration of this process can be found in Fig.~\ref{fig:assessment_flow}.

% problem
For the assessment, the team consisted of a large number of participants from many different disciplines who described the issues they identified in their own language and jargon from their fields. This resulted in a large number of issues, sometimes talking about similar things from slightly different perspectives using different terminology. According to the participants, the large number made manual consolidation as described in \cite{zicari_z-inspection_2021,brusseau_what_2020} both intellectually challenging and labor-intensive. 
They described the main difficulty as identifying which issues could be combined. From reports of previous assessments \cite{brusseau_what_2020,zicari_assessing_2021,zicari_co-design_2021}, it was considered highly likely that different issues could be combined as they describe the same tension, but the final number of tensions, as well as the number of issues per tension, were infeasible to estimate. 

To help in the consolidation process, we decided to use modern text analysis methods to lift semantic meaning from the text based on the concept of Semantic Textual Similarity (STS) \cite{cer_semeval-2017_2017}.
% contribution
In NLP, STS is the task of determining the overlap in meaning between texts. The goal of STS is to provide a numerical score where high values indicate that two texts have similar meanings and low values indicate that their meanings are different \cite{cer_semeval-2017_2017}. 
In this context, the task of identifying issues that describe the same conflict can be seen as identifying and clustering groups of issues that share high STS scores.

We make the following contributions: 
\begin{itemize}[noitemsep,topsep=-1pt]
    \item we show how NLP models can facilitate communication between experts from different domains in trustworthy AI assessment process,
    %\item we give a summary of how deep learning can be used for effective computations of STS,
    \item we present and evaluate two different approaches of STS to group related issues identified by multidisciplinary teams of experts: 1) a clustering-based approach 2) a graph-based approach, both of these use deep learning based STS computation underneath for scoring
    \item we show that the graph-based approach works comparably well to clustering, while not requiring tuning of hyperparameters.
\end{itemize} 
% The novelty of this work is the new graph-based approach that works out of the box without any tuning.
%\textcolor{red}{The novelty of this work are ...}

\section{Method}

\subsection{Word Embeddings and Sentence Embeddings}
Currently, the best performing systems for STS are using deep learning-based embeddings. The basic type of embeddings are word embeddings. In word embeddings, a deep neural network is used to map a word into a fixed dimensional vector space. This mapping is done in a way that captures the meaning of the word so that words with similar meanings have similar vector representations, and analogies in word meanings can be approximated by mathematical operations. As an example, with the analogy ``king is to queen as man is to woman” the encoding $emb_X$ in the vector space should fulfill the equation $emb_{king} - emb_{queen} \approx emb_{man} - emb_{woman}$ \cite{pennington_glove_2014,mikolov_linguistic_2013,mikolov_distributed_2013}.

Sentence embeddings are extensions of word embeddings to complete sentences. Again, deep neural networks are used to map the sentence into a high-dimensional vector space, so that the vector representation also captures the meaning of the sentence \cite{conneau_supervised_2017,cer_universal_2018,reimers_sentence-bert_2019}.  

\subsection{Measuring Semantic Textual Similarity - STS}
After training word or sentence embeddings, the semantic textual similarity of words or sentences is computed from the similarity of their vector representations. A popular metric for this is the cosine similarity. For two words or sentences $A$ and $B$, this is defined as the cosine of the angle $\theta$ between their vector representations $emb_A$ and $emb_B$:
\begin{equation}
    \textit{similarity}(A, B) := cos(\theta) = 
    \frac{emb_A^T \cdot emb_B}{||emb_A|| \cdot ||emb_B||} 
\end{equation}

The computation of STS scores from embeddings is widely used for a variety of tasks such as checking if similar questions were already asked in a forum or the identification of different topics in large text corpora \cite{reimers_sentence-bert_2019}.

\subsection{Identifying groups of similar issues}
For the identification of groups of similar issues, we compared two approaches: 1) clustering-based and 2) graph-based.

\paragraph{Cluster-based group identification.} Separating a set of objects into groups such that objects in the same group have higher similarity and objects in different groups have a lower similarity is the description of a classical clustering problem. Good clustering is best achieved through an iterative process with four key steps: (1) feature selection, (2) cluster identification, (3) cluster validation, and (4) result interpretation. Validation and interpretation are especially important, as algorithms used for cluster identification can always find a division of the objects, but judging whether the division is appropriate and useful, or if a different division should be produced is a decision to be made by the user \cite{xu_survey_2005}.

In our use-case, feature extraction is performed by creating sentence embeddings that map the English text to a high-dimensional vector. An essential strength of this approach is that it allows us to use raw sentences and does not require any preprocessing. This makes the approach straightforward, especially when compared to other approaches where high-quality results may require extensive preprocessing pipelines and tuning \cite{srividhya_evaluating_2010,vijayarani_preprocessing_2015}. The following step is to perform dimensionality reduction, as clustering algorithms are known to have problems when working with high-dimensional vectors. We used UMAP \cite{mcinnes_umap_2020} to map the high-dimensional embedding vectors to lower dimensions, such that most of the relevant local and global structures in the data are preserved \cite{mcinnes_umap_2020}. Compared with other popular dimensionality reduction techniques, UMAP preserves more of the global and local structure of the data than PCA \cite{mcinnes_umap_2020}, while also producing more compact and better separated clusters than t-SNE \cite{mcinnes_umap_2020, kobak_initialization_2021}, which makes it well suited to our task. 

The next step is to iteratively use a clustering algorithm and verify and interpret the resulting clusters until a satisfactory result is found. With this approach, the different clusters correspond to the different groups of issues with high similarity.
Fig.~\ref{fig:clustering_result} in the next section shows the output of this approach.

\paragraph{Graph-based group identification.}
Another approach that works well with data with a similarity measure is spectral clustering \cite{von_luxburg_tutorial_2007}. For spectral clustering, the data is arranged in a weighted, fully connected graph, which is called the similarity graph. In the similarity graph, each node corresponds to a data point and the weight of the edge between two nodes to the similarity of the two associated data points. 
This allows to reformulate the clustering problem into a graph partitioning problem, where the edges between partitions have low weights \cite{von_luxburg_tutorial_2007}.
A popular variation of the similarity graph is the \textit{k-nearest neighbor graph}. With this variation, a node $N_I$ is connected to another node $N_J$, if $N_J$ is among the $k$ nearest neighbors of $N_I$ \cite{von_luxburg_tutorial_2007}.

Applied to the use-case considered here, the nodes in the similarity graph correspond to issues, and the weight of the edge between two nodes corresponds to the cosine similarity of their embeddings.
To simplify the resulting graph, we apply the \textit{1-nearest-neighbor graph} variation, meaning that each node is only connected to the node of it's most similar issue. With this construction, we found that the similarity graph consists of multiple weakly connected components, groups of connected nodes with no connections between nodes from different groups. This simplified the spectral clustering task to identifying the weakly connected components, which in turn provide the separation into groups of issues with high similarity.
In addition, we use the PageRank algorithm \cite{page_pagerank_1999} to assign importance to each of the nodes, based on the connected nodes and their respective importance. The idea behind this is that nodes with many incoming edges are more important and often better better represent an underlying issue compared to nodes with only one incoming edge. 
Fig.~\ref{fig:results_graph} in the next section shows the output of this approach, the outputs of the two approaches will be compared in the next section.
% We show pictorial outputs of both of the above approaches in the next sections.

\section{Experiments} 
In this section, we present the dataset and a subjective evaluation of the results of the two approaches. Code to reproduce our findings is available on GitHub\footnote[1]{\url{https://github.com/dennisrv/iail2022}}.

\subsection{Dataset}
The dataset was made available to us by the authors of the use-case \cite{allahabadi_assessing_2022}, it contains the issues as described by the different expert WGs in a tabular form. Each issue has the following information: an ID, WG name, a title, and a description. The title is a short summary of the issue, while the description provides additional context; the sentence embedding is computed from a concatenation of both. An example issue is listed in Table~\ref{tab:example-issue}.
\bgroup
\begin{table}
    \vspace{-4mm}
    \centering
    \setlength{\tabcolsep}{5pt}
    \renewcommand{\arraystretch}{1.2}
    \caption{Example issue with ID, WG, Title, and Description.}
    \begin{tabularx}{0.9\textwidth}{ | r | X | }
        \hline
        ID & E2 \\
        WG & ethics / healthcare \\ 
        Title & Not all patients may benefit equally from the tool. \\
        Description & The adoption of the system may lead to different care standards for different patient groups. \\
        \hline
    \end{tabularx}
    \label{tab:example-issue}
    % \vspace{-4mm}
\end{table}
\egroup 

In total, the dataset consists of 58 issues described by 51 experts in the six working groups: \textit{technical}, \textit{social}, \textit{ethics}, \textit{ethics / healthcare}, \textit{radiologists}, and \textit{healthcare}. Table~\ref{tab:WGs} gives a summary of size and issues described by each WG.
\bgroup
\begin{table}
    \vspace{-4mm}
    \centering
    \setlength{\tabcolsep}{5pt}
    \renewcommand{\arraystretch}{1.2}
    \caption{Size and number of issues per WG}
    \begin{tabular}{ | c | c | c | }
        \hline
        WG & Members & Issues \\
        \hline
        technical & 21 & 23 \\
        social & 5 & 9 \\ 
        ethics & 3 & 4 \\
        ethics / healthcare & 4 & 8 \\
        radiologists & 3 & 5 \\
        healthcare & 15 & 9 \\
        \hline
        \hline
        total & 51 & 58 \\
        \hline
    \end{tabular}
    \label{tab:WGs}
    \vspace{-4mm}
\end{table}
\egroup

\subsection{Evaluation of different sentence embeddings}
Computations of sentence embeddings are central to our approaches, as this step implicitly defines the similarities between the issues. It is therefore important to use a well-performing NLP model for this task, for which deep neural networks are state-of-the-art \cite{reimers_sentence-bert_2019}. The implementation provided by Reimers et al. \cite{reimers_sentence-bert_2019} makes it possible to use a number of different large, pre-trained networks. For our use case, the \textit{all-mpnet-base-v2} network produced the best results. This network uses MPNet, a transformer architecture with 12 layers, 12 attention heads and a hidden size of 768 \cite{song_mpnet_2020}, which was then fine-tuned for general purpose textual similarity tasks using a dataset with over one billion sentence pairs \cite{reimers_pretrained_nodate}.
In general, we could observe a correlation between the subjective quality of the embeddings and the average performance of the network on several NLP tasks, consistent with the findings in \cite{reimers_pretrained_nodate}. 
With this network, the sentence embeddings are a 768-dimensional real-valued vector.

\subsection{Results of the cluster-based approach}
\begin{figure}[tbph]
    \centering
    \includegraphics[width=0.95\textwidth]{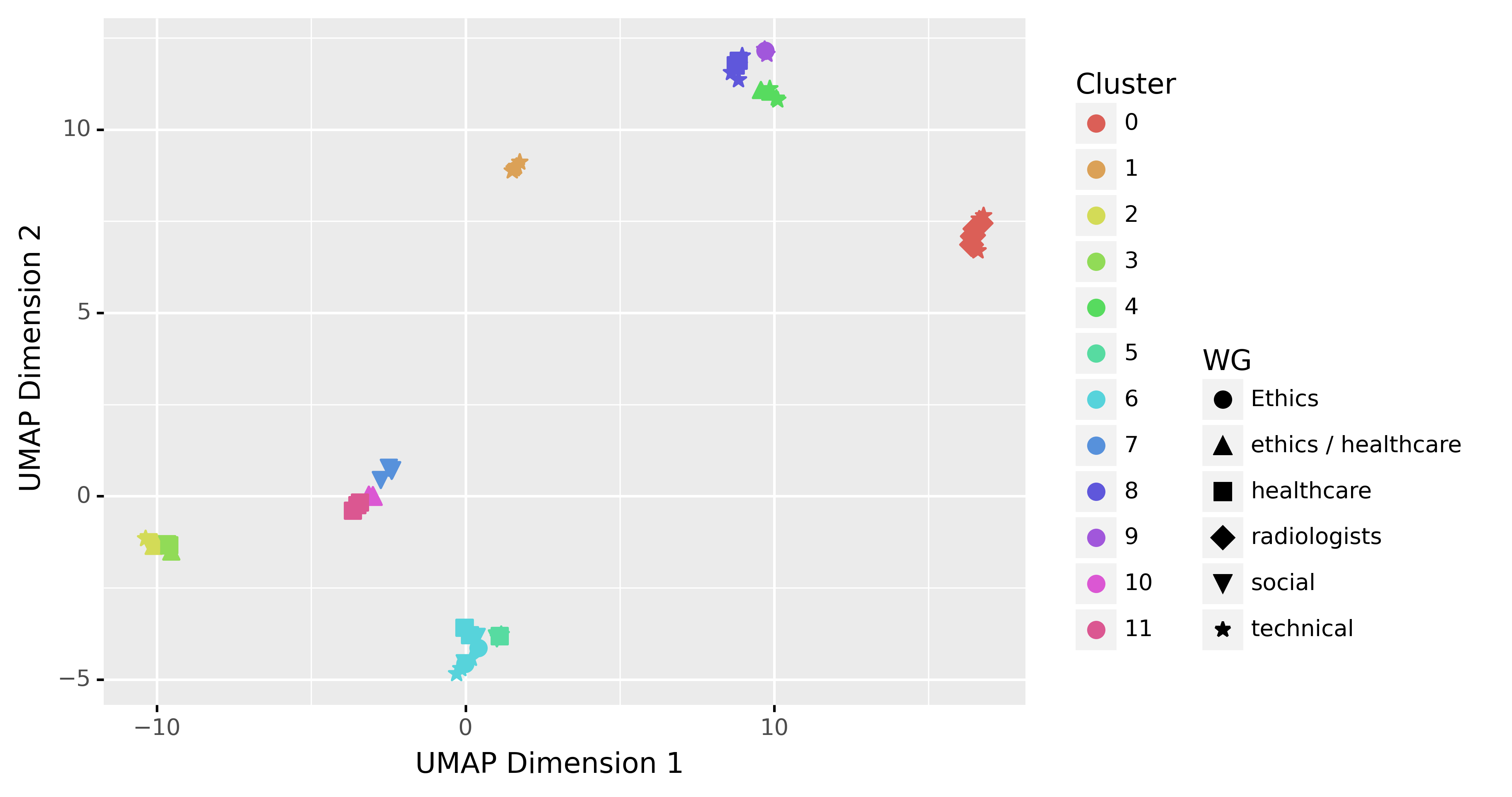}
    \caption{Clustering of issues after using UMAP for a mapping to 2 dimensions that preserves most of the relevant local and global structure in the data.}
    \label{fig:clustering_result}
\end{figure}

The clustering-based approach required some tuning of parameters to achieve a good result. The best results were achieved with a two-step dimensionality reduction with UMAP, which first reduced the 768 dimensions of the sentence embeddings to 15 and then to 2. 
Following this step, we performed a clustering with the HDBSCAN algorithm \cite{mcinnes_hdbscan_2017}, as this  algorithm can find a good number of clusters from the data and does not need the desired number of clusters as an input parameters.
Fig.~\ref{fig:clustering_result} shows the results of the clustering approach.

The result contained 12 groups of issues with most of them containing issues from different WGs. As expected, most of the groups were rather small with 3-5 issues and one larger group containing 9 issues. Through manual inspection, we found that most of the group assignments were reasonable, and only few cases of wrongly assigned issues were found. An example of this is that the issue \textit{Transparency would seem to be enhanced if others could have access to the system} was clustered with issues that were about concerns regarding data safety and privacy.

The strength of this approach is that the low-dimensional mapping with UMAP enables a 2D visualization of the clusters and their relative positions. It is therefore possible to identify cases where a manual inspection could identify that both clusters might be about the same topic, as these clusters will be closer to each other. An example of this are clusters 2 and 3 (bottom left) in Fig.~\ref{fig:clustering_result}. These clusters are thematically related; cluster 2 contains issues about privacy in the dataset, while cluster 3 contains issues about data safety and access to the dataset.

\subsection{Results of the graph-based approach}
Our special construction of the similarity graph and the following simplification of the spectral clustering task to the identification of weakly connected components made it possible to us to omit a pre-specification of the number of clusters, a common input parameter for spectral clustering algorithms \cite{von_luxburg_tutorial_2007}. Instead, the number of clusters emerged naturally as the number of weakly connected components.

\begin{figure}[tpbh]
    \centering
    \includegraphics[width=0.95\textwidth]{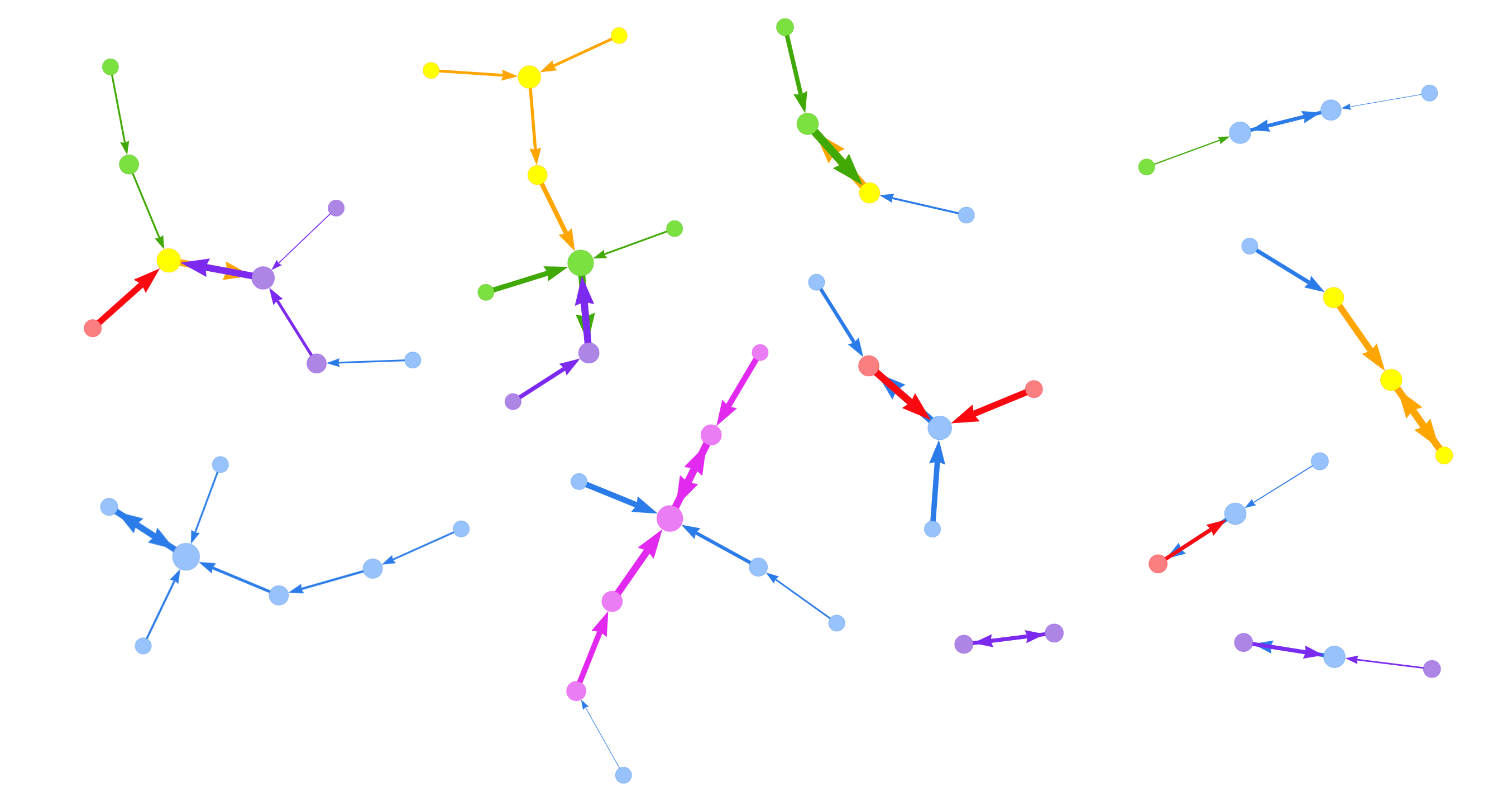}
    \caption{Example of the similarity graph constructed from the issues identified by the experts. The color of a node corresponds to the WG describing the issue. The thickness of the edges is proportional to the similarity of connected issues, the size of nodes is proportional to the importance of the associated issues.}
    \label{fig:results_graph}
\end{figure}

In Fig.~\ref{fig:results_graph} we show the results of the graph-based approach. This approach identified 11 groups of issues (i.e., clusters), with more equally distributed sizes compared to the cluster-based approach. Most of the groups also contained issues from at least 2 different WGs. While the result of this approach and the clustering-based approach were not identical, a manual inspection confirmed that it also produced a reasonable grouping of issues.

The strength of this approach is that it does not require tuning. In addition, nodes with high importance were generally found to capture group content well, which facilitated the manual review. An example of this can be seen later in Table~\ref{tab:overlap_issue} where the top issue is the most important and also captures the problem at a more general level.

\subsection{Comparison of the approaches}
\begin{figure}
    \centering
    \begin{subfigure}[c]{0.48\textwidth}
        \centering
        \includegraphics[width=\textwidth]{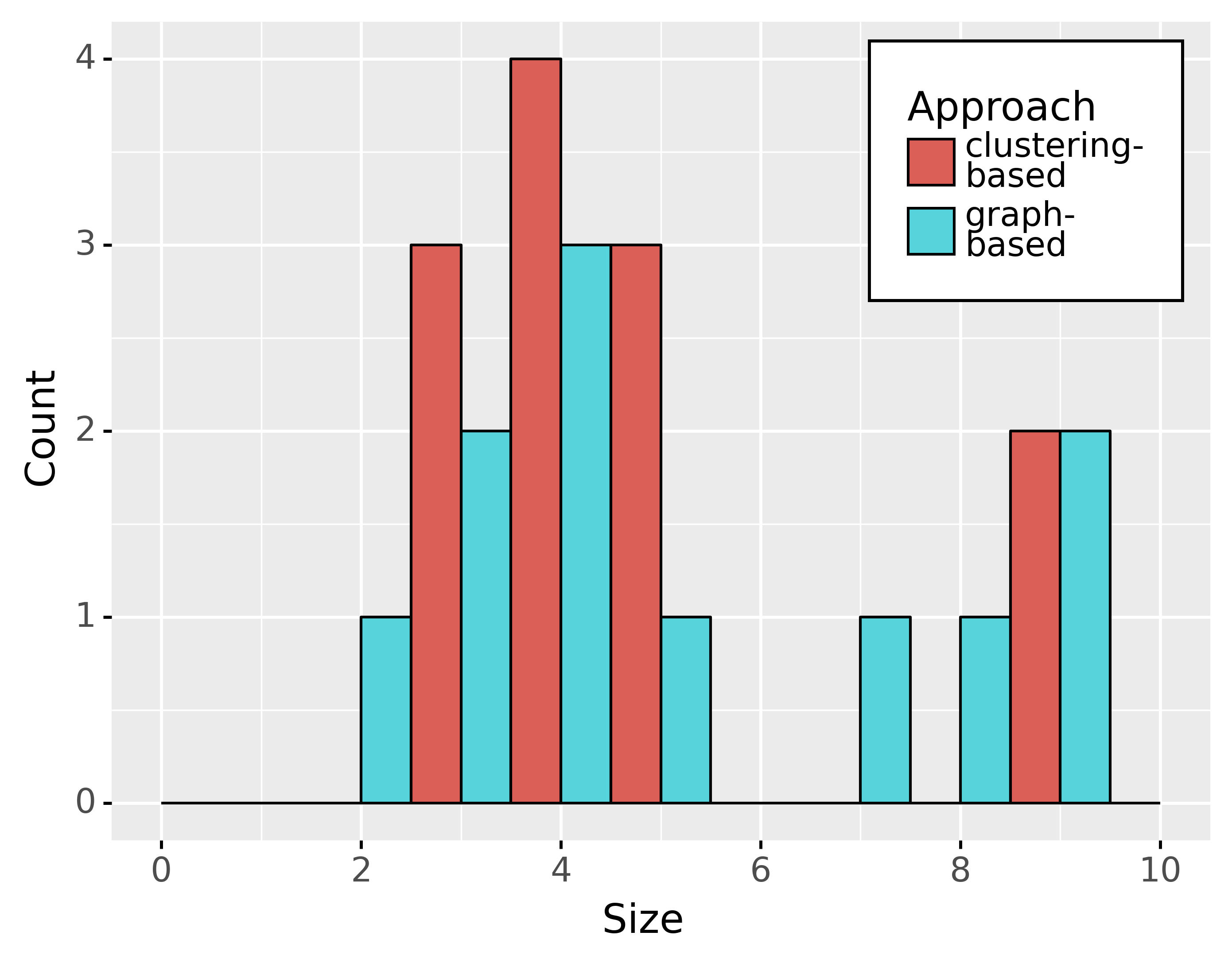}
        \caption{Sizes of groups identified by the different approaches.}
        \label{fig:result_statistics_sizes}
    \end{subfigure}
    \hfill
    \begin{subfigure}[c]{0.48\textwidth}
        \centering
        \includegraphics[width=\textwidth]{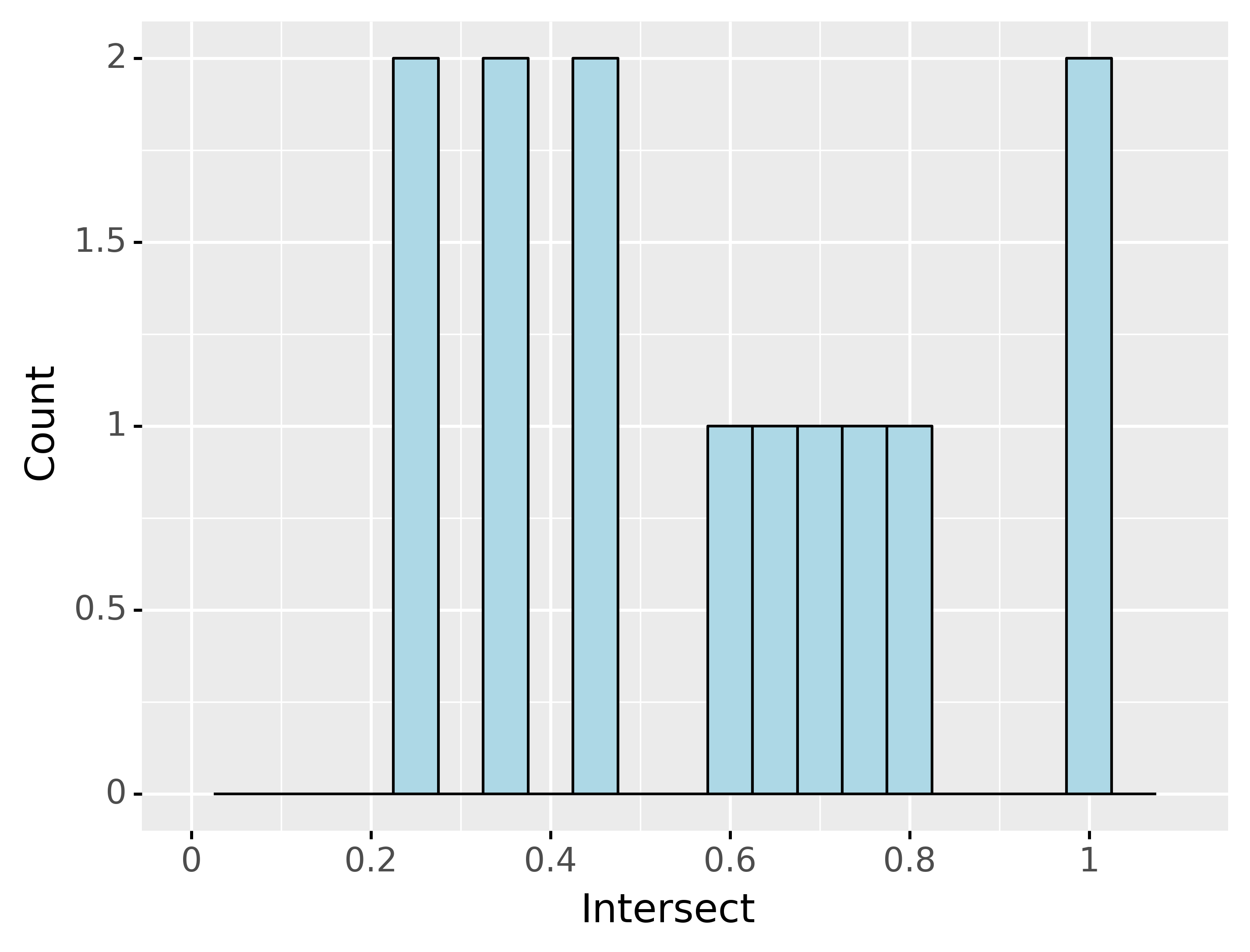}
        \caption{Overlap with the most similar group identified by the other approach.}
        \label{fig:result_statistics_overl}
    \end{subfigure}
    \caption{Histogram of group sizes (a) and overlap with the most similar group identified by the other approach (b).}
    \label{fig:result_statistics}
\end{figure}

Comparing the two approaches, we could observe that the cluster-based approach seemed to prefer grouping the issues in smaller, more specific groups, such as \textit{concerns about stakeholder inclusion} (4 issues) and \textit{concerns on patient benefits} (3 issues). With the graph-based approach, it was found more likely to combine issues from multiple smaller clusters into one larger group, such as \textit{inclusion of and benefit for patients unclear} (9 issues). The differences in size of produced groups are highlighted in Fig.~\ref{fig:result_statistics_sizes}.
Additionally, we found the graph-based approach more likely to assign issues to groups where we could see no clear connection, although with low importance and therefore easy to identify. Contrary, the clustering-based approach subjectively produced less inappropriate groupings, but the lack of an importance within the cluster made the issues that don't belong more difficult to identify.

Furthermore, we could also observe that the two approaches agreed on which issues belonged to the same group in many cases. While often there was no complete agreement, there was still a high overlap between the assigned groups, as shown in Fig.\ref{fig:result_statistics_overl}. For this purpose we computed the overlap of sets of issues $s_A$ and $s_B$ as 
\begin{equation}
    overlap(s_A, s_B) = \frac{| s_A \cap s_B |}{max(|s_A|, |s_B|)}
\end{equation}
An example of a group of issues that both approaches agreed on can be seen in Table~\ref{tab:overlap_issue}.

\bgroup
\begin{table}
    \vspace{-6mm}
    \centering
    \setlength{\tabcolsep}{5pt}
    \renewcommand{\arraystretch}{1.2}
    \caption{Issues that belong to the same group in both approaches. The issues are ordered according to their importance as assigned by the graph-based approach (descending).}
    \begin{tabularx}{0.9\textwidth}{ | c | X | }
        \hline
        WG & Description \\
        \hline
        technical & The dataset used for training is likely not representative for the general population it is currently used on \\
        ethics & [..] there is no way to know whether diverse demographics receive disparate treatment. \\ 
        technical & The model is trained on a particular set of devices and software, undermining the reliability in different scenarios and context. \\
        \hline
    \end{tabularx}
    \label{tab:overlap_issue}
    \vspace{-3mm}
\end{table}
\egroup

For the current assessment \cite{allahabadi_assessing_2022}, we only used the results of the graph-based approach for a pre-screening of the issue groupings during the consolidation phase. However, we plan to use a combination of both clustering approaches for future assessments, as both provide slightly different perspectives and, therefore, are a good start for the discussion between participants. We should also note that in some cases, it was not immediately apparent to the participants whether issues talk about the same problem or not; this could only be solved via discussion and group-consensus.

\subsection{Limitations}
While we found the two approaches to produce sufficient results for our purpose, we could not verify them with data from additional use-cases, as such data was not readily available. 
In addition, we observed cases where the sentence embeddings put too much importance on single words or phrases. For example, the issues \textit{``Transparency would seem to be enhanced if others could have access to the system"} and \textit{``There is a [data safety] concern if data and software engineers have access to the system and others outside of the medical profession"} were assigned to the same group. While the issues have different meanings, the overlap in words used was sufficient to let them appear ``similar enough" to the sentence embedding network. Another occurrence was that all issues containing the word \textit{``Score"} were grouped together, where some of them were later manually assigned to other groups.

Our proposed solution is to have an iterative process in which discussions with the stakeholders about the results of the grouping are conducted, and to use this approach as a support tool only, and not one that gives the definite answer.

\section{Conclusions}
Sentence embeddings and semantic textual similarity can be a useful tool for a Trustworthy AI self-assessment to help an interdisciplinary team of experts and stakeholders with identifying possible risks related to the use of an AI system. 
Our approach was used in practice in a complex use-case with over 50 experts. The approach was used to support initial expert discussions and help build group consensus in a situation where a large number of participants in the assessment made manual consolidation very time-consuming and cumbersome. Participants described it as too demanding for one person to be aware of everyone else's work, making it difficult to find consensus. Instead, our analytical method helped by providing experts with an initial descriptive measure to start the consolidation discussion. 
Since both modeling approaches presented provided an initial result of sufficient and similar quality, we cannot say that one approach is clearly superior to the other. However, the main advantage of both approaches is that they provide an initial grouping of issues. This initial grouping made it much easier to understand the different questions and helped the experts to get a broad picture of the work done by other groups. Because the groupings of questions share a common semantic topic, it was also easier to identify errors in the algorithmic approach and to identify groupings of questions that might belong together. 

To summarize, in the eyes of the participants, the main strength of our method was that it improved their ability to effectively participate in the communication and focus on contributing to the assessment process.
In future assessments, we plan to further validate this approach for consolidation and to investigate with a panel of stakeholders which of the two approaches is more effective for finding consensus.

\section*{}
\subsection*{Funding}
DV received funding from the European Union’s Horizon 2020 research and innovation program under grant agreement no. 101016233 (PERISCOPE), and from the European Union’s Connecting Europe Facility program under grant agreement no. INEA/CEF/ICT/A2020/2276680 (xAIM). The funders had no role in study design, data collection and analysis, decision to publish, or preparation of the manuscript.
\bibliographystyle{splncs04_unsort}
\bibliography{main.bib}

\begin{thebibliography}{10}
\providecommand{\url}[1]{\texttt{#1}}
\providecommand{\urlprefix}{URL }
\providecommand{\doi}[1]{https://doi.org/#1}

\bibitem{whittlestone_ethical_2019}
Whittlestone, J., Nyrup, R., Alexandrova, A., Dihal, K., Cave, S.: Ethical and
  societal implications of algorithms, data, and artificial intelligence: a
  roadmap for research. Nuffield Foundation, London (2019),
  \url{https://www.nuffieldfoundation.org/wp-content/uploads/2019/02/Ethical-and-Societal-Implications-of-Data-and-AI-report-Nuffield-Foundat.pdf}

\bibitem{zicari_z-inspection_2021}
Zicari, R.V., Brodersen, J., Brusseau, J., Düdder, B., Eichhorn, T., Ivanov,
  T., Kararigas, G., Kringen, P., McCullough, M., Möslein, F., Mushtaq, N.,
  Roig, G., Stürtz, N., Tolle, K., Tithi, J.J., van Halem, I., Westerlund, M.:
  Z-{Inspection}®: {A} {Process} to {Assess} {Trustworthy} {AI}. IEEE
  Transactions on Technology and Society  \textbf{2}(2),  83--97 (Jun 2021).
  \doi{10.1109/TTS.2021.3066209}, conference Name: IEEE Transactions on
  Technology and Society

\bibitem{ai_hleg_high-level_expert_group_on_artificial_intelligence_ethics_2019}
on~Artificial~Intelligence, A.H.H.L.E.G.: Ethics guidelines for trustworthy
  {AI}. Text, European Commission (Apr 2019),
  \url{https://op.europa.eu/en/publication-detail/-/publication/d3988569-0434-11ea-8c1f-01aa75ed71a1}

\bibitem{ai_hleg_high-level_expert_group_on_artificial_intelligence_assessment_2020}
on~Artificial~Intelligence, A.H.H.L.E.G.: Assessment {List} for {Trustworthy}
  {Artificial} {Intelligence} ({ALTAI}) for self-assessment. Text, European
  Commission (Jul 2020),
  \url{https://ec.europa.eu/newsroom/dae/document.cfm?doc_id=68342}

\bibitem{signoroni_bs-net_2021}
Signoroni, A., Savardi, M., Benini, S., Adami, N., Leonardi, R., Gibellini, P.,
  Vaccher, F., Ravanelli, M., Borghesi, A., Maroldi, R., Farina, D.:
  {BS}-{Net}: {Learning} {COVID}-19 pneumonia severity on a large chest {X}-ray
  dataset. Medical Image Analysis  \textbf{71},  102046 (Jul 2021).
  \doi{10.1016/j.media.2021.102046},
  \url{https://www.sciencedirect.com/science/article/pii/S136184152100092X}

\bibitem{borghesi_chest_2020}
Borghesi, A., Zigliani, A., Golemi, S., Carapella, N., Maculotti, P., Farina,
  D., Maroldi, R.: Chest {X}-ray severity index as a predictor of in-hospital
  mortality in coronavirus disease 2019: {A} study of 302 patients from
  {Italy}. International Journal of Infectious Diseases  \textbf{96},  291--293
  (Jul 2020). \doi{10.1016/j.ijid.2020.05.021},
  \url{https://linkinghub.elsevier.com/retrieve/pii/S1201971220303283}

\bibitem{leikas_ethical_2019}
Leikas, J., Koivisto, R., Gotcheva, N.: Ethical {Framework} for {Designing}
  {Autonomous} {Intelligent} {Systems}. Journal of Open Innovation: Technology,
  Market, and Complexity  \textbf{5}(1), ~18 (Mar 2019).
  \doi{10.3390/joitmc5010018}, \url{https://www.mdpi.com/2199-8531/5/1/18},
  number: 1 Publisher: Multidisciplinary Digital Publishing Institute

\bibitem{lucivero_ethical_2016}
Lucivero, F.: Ethical {Assessments} of {Emerging} {Technologies}: {Appraising}
  the moral plausibility of technological visions. No.~15 in The
  {International} {Library} of {Ethics}, {Law} and {Technology}, Springer
  International Publishing : Imprint: Springer, Cham, 1st ed. 2016 edn. (2016).
  \doi{10.1007/978-3-319-23282-9}

\bibitem{brusseau_what_2020}
Brusseau, J.: What a {Philosopher} {Learned} at an {AI} {Ethics} {Evaluation}.
  AI Ethics Journal  \textbf{1}(1) (Dec 2020). \doi{10.47289/AIEJ20201214},
  \url{https://www.aiethicsjournal.org/10-47289-aiej20201214}

\bibitem{zicari_assessing_2021}
Zicari, R.V., Brusseau, J., Blomberg, S.N., Christensen, H.C., Coffee, M.,
  Ganapini, M.B., Gerke, S., Gilbert, T.K., Hickman, E., Hildt, E., Holm, S.,
  Kühne, U., Madai, V.I., Osika, W., Spezzatti, A., Schnebel, E., Tithi, J.J.,
  Vetter, D., Westerlund, M., Wurth, R., Amann, J., Antun, V., Beretta, V.,
  Bruneault, F., Campano, E., Düdder, B., Gallucci, A., Goffi, E., Haase,
  C.B., Hagendorff, T., Kringen, P., Möslein, F., Ottenheimer, D., Ozols, M.,
  Palazzani, L., Petrin, M., Tafur, K., Tørresen, J., Volland, H., Kararigas,
  G.: On {Assessing} {Trustworthy} {AI} in {Healthcare}. {Machine} {Learning}
  as a {Supportive} {Tool} to {Recognize} {Cardiac} {Arrest} in {Emergency}
  {Calls}. Frontiers in Human Dynamics  \textbf{3}, ~30 (2021).
  \doi{10.3389/fhumd.2021.673104},
  \url{https://www.frontiersin.org/article/10.3389/fhumd.2021.673104}

\bibitem{zicari_co-design_2021}
Zicari, R.V., Ahmed, S., Amann, J., Braun, S.A., Brodersen, J., Bruneault, F.,
  Brusseau, J., Campano, E., Coffee, M., Dengel, A., Düdder, B., Gallucci, A.,
  Gilbert, T.K., Gottfrois, P., Goffi, E., Haase, C.B., Hagendorff, T.,
  Hickman, E., Hildt, E., Holm, S., Kringen, P., Kühne, U., Lucieri, A.,
  Madai, V.I., Moreno-Sánchez, P.A., Medlicott, O., Ozols, M., Schnebel, E.,
  Spezzatti, A., Tithi, J.J., Umbrello, S., Vetter, D., Volland, H.,
  Westerlund, M., Wurth, R.: Co-{Design} of a {Trustworthy} {AI} {System} in
  {Healthcare}: {Deep} {Learning} {Based} {Skin} {Lesion} {Classifier}.
  Frontiers in Human Dynamics  \textbf{3}, ~40 (2021).
  \doi{10.3389/fhumd.2021.688152},
  \url{https://www.frontiersin.org/article/10.3389/fhumd.2021.688152}

\bibitem{cer_semeval-2017_2017}
Cer, D., Diab, M., Agirre, E., Lopez-Gazpio, I., Specia, L.: {SemEval}-2017
  {Task} 1: {Semantic} {Textual} {Similarity} {Multilingual} and {Crosslingual}
  {Focused} {Evaluation}. In: Proceedings of the 11th {International}
  {Workshop} on {Semantic} {Evaluation} ({SemEval}-2017). pp. 1--14.
  Association for Computational Linguistics, Vancouver, Canada (2017).
  \doi{10.18653/v1/S17-2001}, \url{http://aclweb.org/anthology/S17-2001}

\bibitem{pennington_glove_2014}
Pennington, J., Socher, R., Manning, C.: Glove: {Global} {Vectors} for {Word}
  {Representation}. In: Proceedings of the 2014 {Conference} on {Empirical}
  {Methods} in {Natural} {Language} {Processing} ({EMNLP}). pp. 1532--1543.
  Association for Computational Linguistics, Doha, Qatar (2014).
  \doi{10.3115/v1/D14-1162}, \url{http://aclweb.org/anthology/D14-1162}

\bibitem{mikolov_linguistic_2013}
Mikolov, T., Yih, W.t., Zweig, G.: Linguistic regularities in continuous space
  word representations. In: Proceedings of the 2013 conference of the north
  american chapter of the association for computational linguistics: {Human}
  language technologies. pp. 746--751 (2013)

\bibitem{mikolov_distributed_2013}
Mikolov, T., Sutskever, I., Chen, K., Corrado, G.S., Dean, J.: Distributed
  {Representations} of {Words} and {Phrases} and their {Compositionality}. In:
  Advances in {Neural} {Information} {Processing} {Systems}. vol.~26. Curran
  Associates, Inc. (2013),
  \url{https://proceedings.neurips.cc/paper/2013/hash/9aa42b31882ec039965f3c4923ce901b-Abstract.html}

\bibitem{conneau_supervised_2017}
Conneau, A., Kiela, D., Schwenk, H., Barrault, L., Bordes, A.: Supervised
  {Learning} of {Universal} {Sentence} {Representations} from {Natural}
  {Language} {Inference} {Data}. In: Proceedings of the 2017 {Conference} on
  {Empirical} {Methods} in {Natural} {Language} {Processing}. pp. 670--680.
  Association for Computational Linguistics, Copenhagen, Denmark (Sep 2017).
  \doi{10.18653/v1/D17-1070}, \url{https://aclanthology.org/D17-1070}

\bibitem{cer_universal_2018}
Cer, D., Yang, Y., Kong, S.y., Hua, N., Limtiaco, N., John, R.S., Constant, N.,
  Guajardo-Cespedes, M., Yuan, S., Tar, C., Sung, Y.H., Strope, B., Kurzweil,
  R.: Universal {Sentence} {Encoder}. arXiv:1803.11175 [cs]  (Apr 2018),
  \url{http://arxiv.org/abs/1803.11175}, arXiv: 1803.11175

\bibitem{reimers_sentence-bert_2019}
Reimers, N., Gurevych, I.: Sentence-{BERT}: {Sentence} {Embeddings} using
  {Siamese} {BERT}-{Networks}. arXiv:1908.10084 [cs]  (Aug 2019),
  \url{http://arxiv.org/abs/1908.10084}, arXiv: 1908.10084

\bibitem{xu_survey_2005}
Xu, R., Wunsch, D.C.: Survey of {Clustering} {Algorithms}. IEEE TRANSACTIONS ON
  NEURAL NETWORKS  \textbf{16}(3), ~35 (2005)

\bibitem{srividhya_evaluating_2010}
Srividhya, V., Anitha, R.: Evaluating preprocessing techniques in text
  categorization. International journal of computer science and application
  \textbf{47}(11),  49--51 (2010),
  \url{http://sinhgad.edu/ijcsa-2012/pdfpapers/1_11.pdf}

\bibitem{vijayarani_preprocessing_2015}
Vijayarani, D.S., Ilamathi, J., Nithya: Preprocessing {Techniques} for {Text}
  {Mining} - {An} {Overview}. International Journal of Computer Science \&
  Communication Networks  \textbf{5}, ~11 (2015)

\bibitem{mcinnes_umap_2020}
McInnes, L., Healy, J., Melville, J.: {UMAP}: {Uniform} {Manifold}
  {Approximation} and {Projection} for {Dimension} {Reduction}.
  arXiv:1802.03426 [cs, stat]  (Sep 2020),
  \url{http://arxiv.org/abs/1802.03426}, arXiv: 1802.03426

\bibitem{kobak_initialization_2021}
Kobak, D., Linderman, G.C.: Initialization is critical for preserving global
  data structure in both t-{SNE} and {UMAP}. Nature Biotechnology
  \textbf{39}(2),  156--157 (Feb 2021). \doi{10.1038/s41587-020-00809-z},
  \url{https://www.nature.com/articles/s41587-020-00809-z}, number: 2
  Publisher: Nature Publishing Group

\bibitem{von_luxburg_tutorial_2007}
von Luxburg, U.: A tutorial on spectral clustering. Statistics and Computing
  \textbf{17}(4),  395--416 (Dec 2007). \doi{10.1007/s11222-007-9033-z},
  \url{http://link.springer.com/10.1007/s11222-007-9033-z}

\bibitem{page_pagerank_1999}
Page, L., Brin, S., Motwani, R., Winograd, T.: The {PageRank} {Citation}
  {Ranking}: {Bringing} {Order} to the {Web}. Technical {Report} 1999-66,
  Stanford InfoLab (Nov 1999), \url{http://ilpubs.stanford.edu:8090/422/},
  backup Publisher: Stanford InfoLab

\bibitem{allahabadi_assessing_2022}
Allahabadi, H., Amann, J., Balot, I., Beretta, A., Binkley, C., Bozenhard, J.,
  Bruneault, F., Brusseau, J., Candemir, S., Cappellini, L.A., Castagnet, G.F.,
  Chakraborty, S., Cherciu, N., Cociancig, C., Coffee, M., Ek, I.,
  Espinosa-Leal, L., Farina, D., Fieux-Castagnet, G., Frauenfelder, T.,
  Gallucci, A., Giuliani, G., Golda, A., van Halem, I., Hildt, E., Holm, S.,
  Kararigas, G., Krier, S.A., Kühne, U., Lizzi, F., Madai, V.I., Markus, A.F.,
  Masis, S., Mathez, E.W., Mureddu, F., Neri, E., Osika, W., Ozols, M.,
  Panigutti, C., Parent, B., Pratesi, F., Moreno-Sánchez, P.A., Sartor, G.,
  Savardi, M., Signoroni, A., Sormunen, H., Spezzatti, A., Srivastava, A.,
  Stephansen, A.F., Theng, L.B., Tithi, J.J., Tuominen, J., Umbrello, S.,
  Vaccher, F., Vetter, D., Westerlund, M., Wurth, R., Zicari, R.V.: Assessing
  {Trustworthy} {AI} in times of {COVID}-19. {Deep} {Learning} for predicting a
  multi-regional score conveying the degree of lung compromise in {COVID}-19
  patients. IEEE Transactions on Technology and Society pp.~1--1 (2022).
  \doi{10.1109/TTS.2022.3195114}

\bibitem{song_mpnet_2020}
Song, K., Tan, X., Qin, T., Lu, J., Liu, T.Y.: {MPNet}: {Masked} and {Permuted}
  {Pre}-training for {Language} {Understanding}. arXiv:2004.09297 [cs]  (Nov
  2020), \url{http://arxiv.org/abs/2004.09297}, arXiv: 2004.09297

\bibitem{reimers_pretrained_nodate}
Reimers, N.: Pretrained {Models} — {Sentence}-{Transformers} documentation,
  \url{https://www.sbert.net/docs/pretrained_models.html}

\bibitem{mcinnes_hdbscan_2017}
McInnes, L., Healy, J., Astels, S.: hdbscan: {Hierarchical} density based
  clustering. The Journal of Open Source Software  \textbf{2}(11), ~205 (Mar
  2017). \doi{10.21105/joss.00205},
  \url{http://joss.theoj.org/papers/10.21105/joss.00205}

\end{thebibliography}

\end{document}